# SPECTROSCOPIC STUDIES ON NANOCOMPOSITES OBTAINED BY FUNCTIONALIZATION OF CARBON NAOTUBES WITH CONDUCTING POLYMERS

Serge Lefrant[1*], Mihaela Baibarac[2] and Ioan Baltog[2]

[1]Institut des Matériaux Jean Rouxel, 2 rue de la Houssinière,

B.P. 32229, 44322 Nantes, Cedes 03, France

[2] National Institute of Materials Physics, Lab. Optics and Spectroscopy,

Bucharest, P.O.Box MG-7, R-76900, Romania

**ABSTRACT**

Vibrational properties of composites based on single-walled carbon nanotubes (SWNTs) and conducting polymers of the type polyaniline (PANI) and poly (3,4-ethylene dioxythiophene) (PEDOT) are reported. For PANI-functionalized SWNTs, the intensity increase of the Raman band at 178 cm$^{-1}$, associated with radial breathing modes of SWNTs bundles, indicates an additional roping of nanotubes due to the presence of the polymer. The interaction of this composite with $NH_4OH$ solution involves an internal redox reaction between PANI and SWNTs. Thus, the polymer chain undergoes a transition from the semi-oxidized state into a reduced one. The functionalization of SWNT side walls with PEDOT is invoked as well.

*Corresponding author: Prof. S. Lefrant

Tel: +33 240 373 910,

Fax: +33 240 373 995,

E-mail: Serge.Lefrant@cnrs-imn.fr

## 1. INTRODUCTION

Many efforts have been made to combine carbon nanotubes (CNTs) and polymers to produce functional composite materials with improved properties.[1-3] The detailed investigation of conducting polymer/carbon nanotubes (CP/CNTs) composites is nowadays of great interest. These materials have applications in different fields such as supercapacitors, sensors, photovoltaic cells and photodiodes, optical limiting devices, solar cells, high resolution printable conductors, electromagnetic absorbers and last but not least advanced transistors.

The preparation of composites requires to qualify the type of the interaction between the host matrix and the guest nanoparticles. In CP/CNTs composites, it has been shown that either the polymer functionalizes CNTs[4,5] or the CPs are doped with CNTs, when a charge transfer between the two constituents takes place[5-7]. Three routes are used to prepare CP/CNTs composites: direct mixing of the CP with CNTs, chemical synthesis of the CP in the presence of CNTs and electropolymerization of monomers on the CNTs film. In this paper, we show by Raman and FTIR spectroscopy that the last method is the most efficient route to achieve the functionalization of SWNTs with CPs.

## 2. EXPERIMENTAL

PANI and PEDOT were electrochemically prepared via cyclic voltammetry in a three-electrode cell containing an aqueous solution of: i) 0.5 M HCl and 0.05 M aniline (AN) and ii) 0.002 M benzyl dimethyl hexadecylammonium chloride (BDHAC) and 0.05M 3,4-ethylene dioxythiophene (EDOT). SWNT films deposited on a 25 $mm^2$ Au plate were used as working electrodes and a Pt spiral as counter electrode. The reference electrode was a saturated calomel electrode (SCE). PANI and PEDOT films were obtained by cycling in the potential ranges (-200; +700) and (-800; +800) mV vs. SCE, respectively at a sweep rate of 100 mV $s^{-1}$. The de-doping of the PANI-ES film was achieved with $NH_4OH$.

Electrochemical measurements were carried out using a potentiostat / galvanostat type Princeton Applied Research (PAR), model 173, a PAR pulse generator, model 175 and a Philips-type X-Y recorder. SERS studies were performed in a backscattering geometry, under laser excitation wavelengths 1064 and 676.4 nm, using a RFS 100 FT Raman Bruker spectrophotometer and a Jobin Yvon T64000 Raman spectrophotometer, respectively. FTIR spectra were acquired with a Bruker IFS 28 spectrophotometer.

## 3. RESULTS AND DISCUSSION

Fig. 1 shows cyclic votammetry curves recorded on a Au electrode coated with a SWNT film immersed into a 0.5 M HCl solution and 0.05 M aniline in the potential range (-200; +800) mV vs. SCE. As the successive number of scans increases, a gradual shift of both oxidation and reduction peaks is observed, indicating a lower kinetics of electron transfer. Besides, for the first 25 cyclic voltammetry curves, a linear proportionality between the oxidation and reduction peak current and the scan rate was found, showing characteristics of thin layer electrochemical behavior. The presence of the thin layer of PANI on the CNT film is well put en evidence in Fig. 2, curve 2. The Raman spectrum of the SWNT film is shown in Fig. 2, curve 1. In the low frequency range below 300 $cm^{-1}$, one finds the bands associated with the radial breathing modes (RBM). The intensity and peak position of these bands, related to the tube diameter through the relation $\nu$ ($cm^{-1}$) = 223.75/d (nm)[8], are very sensitive to the excitation wavelength. The second group, consisting of the G and D bands, is found in the interval from 1100 to 1700 $cm^{-1}$. Both bands are not only related to the nanotubes structure; the former, situated at 1595 $cm^{-1}$ attributed to in-plane $E_{2g}$ vibration mode, is also present in the Raman spectrum of other graphitic materials.[8] Regarding the D band, it is considered as an indication of disorder in graphite lattices or defects in nanotubes.[8] In accordance with spectra 2-4 and 6-8 from Fig.2, increasing the number of cycles from 25 to 300 modifies significantly the Raman spectrum of SWNTs as it follows: i) the Raman features of PANI-emeraldine salt (ES) appear and increase in intensity when the number of cycles grows. In the 1100-1700 $cm^{-1}$ spectral range, the Raman lines of PANI-ES are situated at ca. 1175, 1330-1380, 1506, 1589 and 1618 $cm^{-1}$. They are attributed to the following vibrational modes: C-H bending in benzenoid rings (B) – $A_g$ mode, semiquinone radical structure, C=N stretching, C=C stretching in quinoid rings (Q) and C-C stretching in benzenoid rings (B), respectively[9]; ii) in the low-frequency region, increasing the number of cycles from 25 to 75 (curves 2 to 4, upper left panel) leads to a sudden decrease of the RBM of isolated tubes (164 $cm^{-1}$) whereas the Raman line associated with bundled tubes (178 $cm^{-1}$)[10] increases gradually. This suggests that the relative proportion of isolated tubes decreases, because additional roping is induced by the electrochemical polymerization. The increase of the cycles number from 100 to ca. 300 (curves 6 – 8, Fig.2) leads to a progressive decrease in intensity of the Raman line situated at 178 $cm^{-1}$ until its complete disappearance. After 300 cycles (curve 8, Fig. 2), the Raman spectrum is fully dominated by the spectral features of PANI-ES and the complex band with maximum at ca. 85 - 118 $cm^{-1}$. A puzzling fact is that the same band at 85-118 $cm^{-1}$ has also been observed on composites prepared by direct mixing of SWNTs with the PANI-EB solutions.[6] Though it is known that the interaction of PANI-salt with a base leads to the PANI-base form, by the treatment of the PANI-salt functionalized SWNTs films, obtained after 75 and 300 cycles (Fig.2, spectra 5 and 9), with a $NH_4OH$ 1M solution, it is easily observed that the Raman lines of SWNTs at 178, 1277 and 1595 $cm^{-1}$ are quite fully restored while the PANI lines are almost totally absent, whatever the number of cycles. At a first sight, the effect of the $NH_4OH$ treatment on the PANI-salt functionalized SWNTs films is difficult to understand. In Fig. 3 the presence of Raman lines at 1156, 1216 and 1464 $cm^{-1}$, all belonging to PANI, confirms that the interaction of PANI-salt functionalized SWNTs film with $NH_4OH$ cannot be regarded as a reaction of chemical defunctionalization of SWNTs. Conversely, it remains the possibility that the composite post-treated with $NH_4OH$ contains PANI in the LB state which is difficult to detect by Raman scattering under a 1064 nm laser excitation.[9] In our opinion, this hypothesis may explain the absence of the PANI Raman lines on curves 5 and 9 from Fig. 1.

More information concerning both the oxidation degree of PANI chemically bonded on the SWNT surface and the functionalization mechanism is offered of the FTIR spectra from Fig. 4. Two new absorption bands can be seen on curve 4 from Fig. 4a. One is situated at ~770 $cm^{-1}$ and the other at ~740 $cm^{-1}$. The two bands are due to deformation vibrations of the benzenoid and quinoid ring, respectively.[9] Their appearance indicate the formation of two types of composites during the electropolymerization process:

A

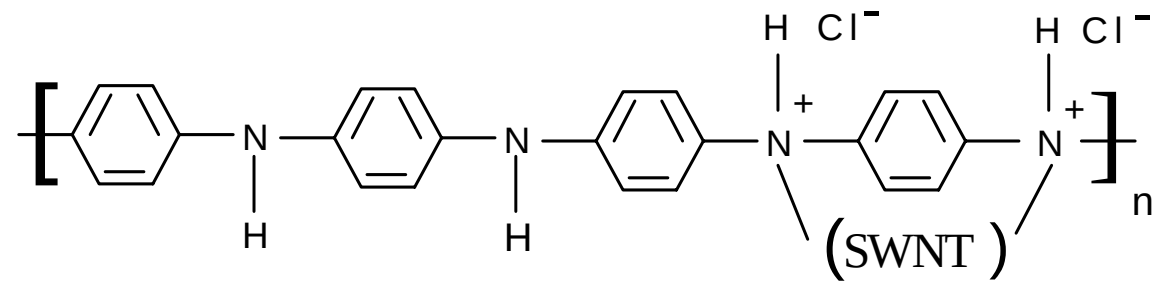

B

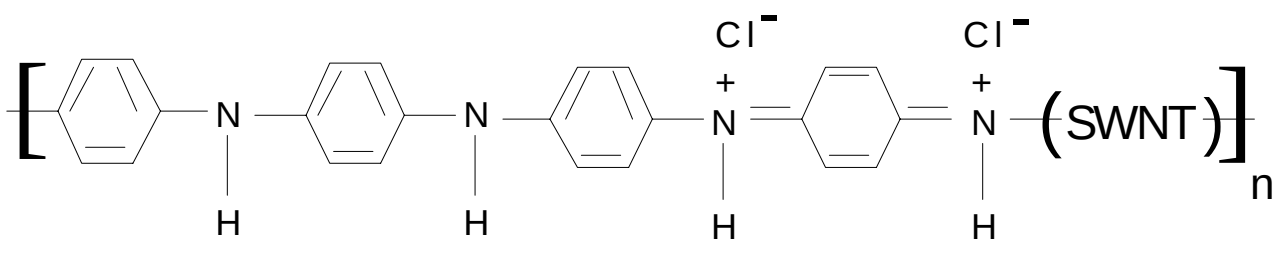

The transformation the PANI-salt functionalized SWNTs in PANI-base functionalized SWNTs (curves 4 in Fig. 4, a and b) involves: i) the modification of the ratio between the FTIR bands intensity at ca. 770 and 750 $cm^{-1}$; ii) the increase in the intensity of the 1219 $cm^{-1}$ band which is attributed to C-N stretching + ring deformation (B) + C-H bending (B) vibration mode[9]; iii) the appearance of new bands at ca. 1458, 1540 and 1559 $cm^{-1}$ attributed to C-C stretching, C-H and N-H bending vibration[9] and iv) the disappearance of the complex band in the 1560-1625 $cm^{-1}$ spectral range. A $NH_4OH$ post-treatment of the PANI-salt functionalized SWNT film indicates that a supplementary functionalization of SWNTs with the polymer takes place. In this case, we invoke an internal redox reaction between PANI-EB and SWNTs which results in the transition of the polymer from a semi-oxidized state into a reduced one. We note that the reaction of A and B compounds with $NH_4OH$ leads to the formation of PANI-LB functionalized SWNTs and PANI-EB functionalized SWNTs, respectively. In this way, new covalent C-N bonds are formed between the imine nitrogen atoms of the repeating units of PANI-EB and the carbon atoms of SWNTs. It is known that the intensity ratio of the bands at ca. 1592 and 1500 $cm^{-1}$ can provide information on the degree of oxidation of the polymer.[9] This ratio, in the case of the curve 4 from Fig. 4b, indicates that leucoemeraldine base-functionalized SWNTs are formed. The enhancement of the bands at 740-750 and 772 $cm^{-1}$ for PANI-functionalized SWNT composites (Fig.4, a and b) is quite explainable by strong hindrance effects induced by the binding of SWNTs as whole units on the polymer chain. The redox reaction between PANI-EB and SWNTs takes place with a smaller efficiency when the number of cycles increases, this fact being put in evidence on curves 2 and 3 from Fig. 4b, where FTIR spectra contain also some spectral features of PANI-EB functionalized SWNTs composite. In this context, coming back to spectrum 9 of Fig. 2, the small band at 1165 $cm^{-1}$ is not a speculative spectral detail since it confirms the presence of PANI-EB with low weight in the type PANI-base covalently functionalized SWNTs composite.

Raman spectra recorded after electropolymerization of EDOT on the SWNTs film (Fig. 5) show: i) in the 50-250 $cm^{-1}$ spectral range, a similarly variation with that reported in the case of electropolymerization of aniline on the SWNT film. An additional fact is the up-shift of the RBM Raman line at 164 and 178 $cm^{-1}$ associated with individual and bundled tubes, respectively[11]; ii) significant variations in the range 900-1780 $cm^{-1}$, which concern both PEDOT and SWNTs. In Fig.5 the main Raman lines of PEDOT are situated to: 943 - 988, 1100-1136, 1263, 1364, 1433 and 1532 $cm^{-1}$. They are attributed to the following vibrational modes: oxyethylene ring deformation, C-O-C deformation, $C_\alpha$ - $C_{\alpha'}$ stretching + C-H bending, $C_\beta$-$C_{\beta'}$ stretching, symmetric C=C stretching and asymmetric C=C stretching, respectively.[11] The Raman features of PEDOT appear and increase in intensity when the number of cycles grows. Besides, for PEDOT it is observed: i) a change of the intensity ratio of Raman lines situated in the 900-1000 $cm^{-1}$ spectral range; ii) a down-shift of the Raman line peaks to 1100 $cm^{-1}$; iii) an up-shift of Raman lines from 1136 $cm^{-1}$; iv) an increase in the intensity of the Raman lines with maximum at 1453 $cm^{-1}$. Concerning SWNTs, we note that the G band consists of four Raman lines at ca. 1555, 1573, 1595 and 1610 $cm^{-1}$.[8,12] The increase of the cycle number from 100 to ca. 300 (curves 3-6, Fig.5) leads to: i) a progressive increase in the relative intensity of the Raman line situated at 1570 $cm^{-1}$. After 300 cycles (curve 6, Fig. 5), the intensity ratio of Raman lines from 1570 and 1594 $cm^{-1}$ is equal to 1; ii) an up-shift of the D band from 1275 to 1300 $cm^{-1}$ and iii) the appearance of a new Raman band in the spectrum range 1600-1750 $cm^{-1}$. We think that the modifications mentioned above can be explained on the base of a functionalization of SWNTs side-walls with PEDOT. This assumption is supported by Raman data recently reported on SWNTs.[13,14] Generally, the functionalization of SWNTs involves a gradual increase of defects on nanotubes. They induce an up-shift in the RBM region[13] and an increase in the relative intensity of the lines at 1555 and 1525 $cm^{-1}$, the later being assigned to an $E_3$ symmetry disorder-induced peak,[14] features also presented in Fig. 5. Taking all these results into account, significant changes of the Raman lines which compose the G band suggest a functionalization of SWNTs side-walls with the polymer. This can also explain variations of the spectral features of PEDOT from the 943 – 988 and 1100-1136 $cm^{-1}$ spectral ranges which originates in the strong steric hindrance effects induced by the functionalization of SWNTs with polymer.

## 4. CONCLUSIONS

The main results are summarized as follows: i) the covalent functionalization of SWNTs with PANI is obtained in two successive stages. The first one corresponds to the electro-polymerization of aniline in an HCl solution on a SWNT film which results in composites on the type PANI-LS functionalized SWNTs and PANI-ES functionalized SWNTs. The second one is the result of $NH_4OH$ post treatment on PANI-(LS or ES) functionalized SWNTs; ii) the increase in the intensity of the Raman band at 178 $cm^{-1}$ during the electropolymerization of aniline on CNs film indicates an additional nanotube roping with PANI as a binding agent; iii) the binding of SWNTs as whole units on the PANI chain induces strong steric hindrance effects observed in FTIR spectra by bands at 740 - 750 and 772 $cm^{-1}$; iv) the electropolymerization of EDOT on the SWNTs film results in a functionalization of the side-walls of nanotubes with the polymer.

**ACKNOWLEDGMENTS**

This work was performed in the frame of the Scientific Cooperation between the Laboratory of Crystalline Physics of the Institute of Materials Jean Rouxel in Nantes, and the Laboratory of Optics and Spectroscopy of the National Institute of Materials Physics, Bucharest.

## Figure Caption

**Fig. 1** Cyclic voltammograms on an Au support coated with a SWNT film carried out in 0.5 M HCl aqueous solution and aniline 0.05 M in potential range (-200; +800) mV vs. SCE, recorded with a sweep rate of 50 mV $s^{-1}$ – cycles: 1, 10 and 25 (a). Dependence of anodic (b) and cathodic (c) peak current at 200 mV vs. SCE of the sweep rate.

**Fig. 2** Raman spectra at $\lambda_{exc}$ = 1064 nm of PANI/SWNTs composites obtained by electro-polymerization of AN on a SWNTs film in HCl 0.5M. Curve 1 corresponds to the SWNTs film Raman spectrum. Curves 2-4 and 6-8 show the evolution of the Raman spectrum after 25, 50, 75, 100, 150 and 300 cycles, respectively carried out in the potential range (-200; +700) mV vs. SCE with a sweep rate of 100 mV $s^{-1}$. De-doping of the PANI-salt functionalized SWNTs films (curves 4 and 8), as a result of the chemical reaction with the NH4OH 1M solution, is illustrated on curves 5 and 9.

**Fig. 3** Raman spectra at $\lambda_{exc}$ = 676.4 nm of: a) SWNTs films of ca. 350 nm thickness deposited on rough Au support and b) PANI-salt functionalized SWNTs composite electrochemically prepared using HCl 0.5M after ca. 75 cycles carried out in the potential range (-200; +700) mV vs. SCE with a sweep rate of 100 mV $s^{-1}$ and followed by $NH_4OH$ 1M post treatment.

**Fig. 4** FTIR spectra of PANI-ES (curve 1, a) and PANI salt-functionalized SWNTs composites (curves 2 - 4, a). These composites were electrochemically prepared by the achievement of 75, 150 and 300 cycles (curves 4, 3 and 2, respectively) on the SWNTs film, immersed in the solution of AN and HCl 0.5M, in the potential range (-200; +700) mV vs. SCE with a sweep rate of 100 mV $s^{-1}$. FTIR spectra of PANI-EB (curve 1, b) and PANI-base functionalized SWNTs composites (curves 2 - 4, b) obtained by a subsequent reaction of PANI-salt functionalized SWNTs (curves 2 - 4, a) with an $NH_4OH$ 1M solution.

**Fig. 5** Raman spectra at $\lambda_{exc}$ = 1064 nm of PEDOT /SWNTs composites obtained by electro-polymerization of EDOT on a SWNT film in BDHAC solution. Curve 1 corresponds to the SWNTs film Raman spectrum. Curves 2-6 show the evolution of the Raman spectrum after 50, 100, 150, 200 and 300 cycles, respectively carried out in the potential range (-800; +800) mV vs. SCE with a sweep rate of 100 mV $s^{-1}$. Raman spectra of the PEDOT film deposited on Au support in same conditions after 200 cycles.

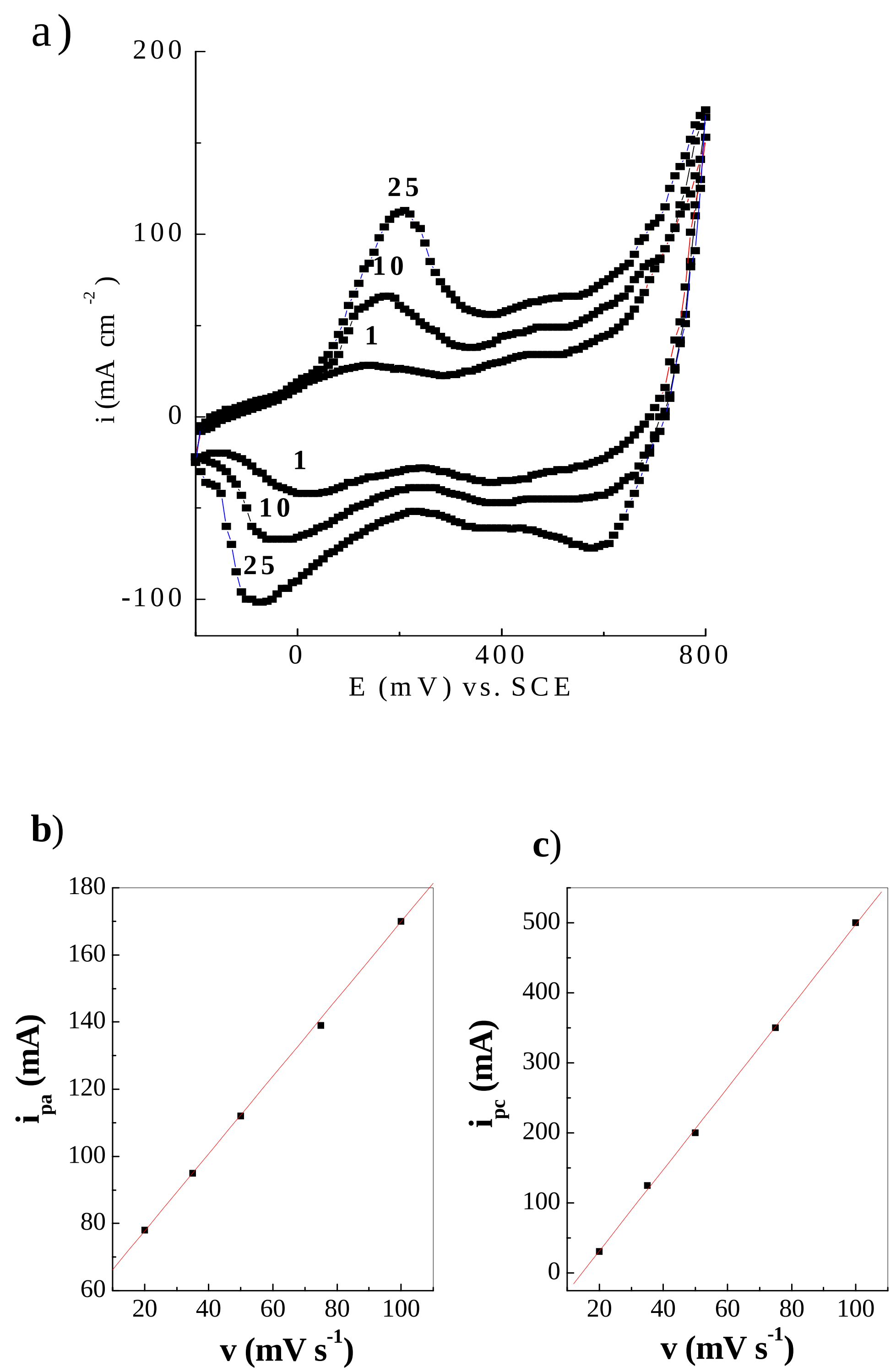

**Fig. 1**

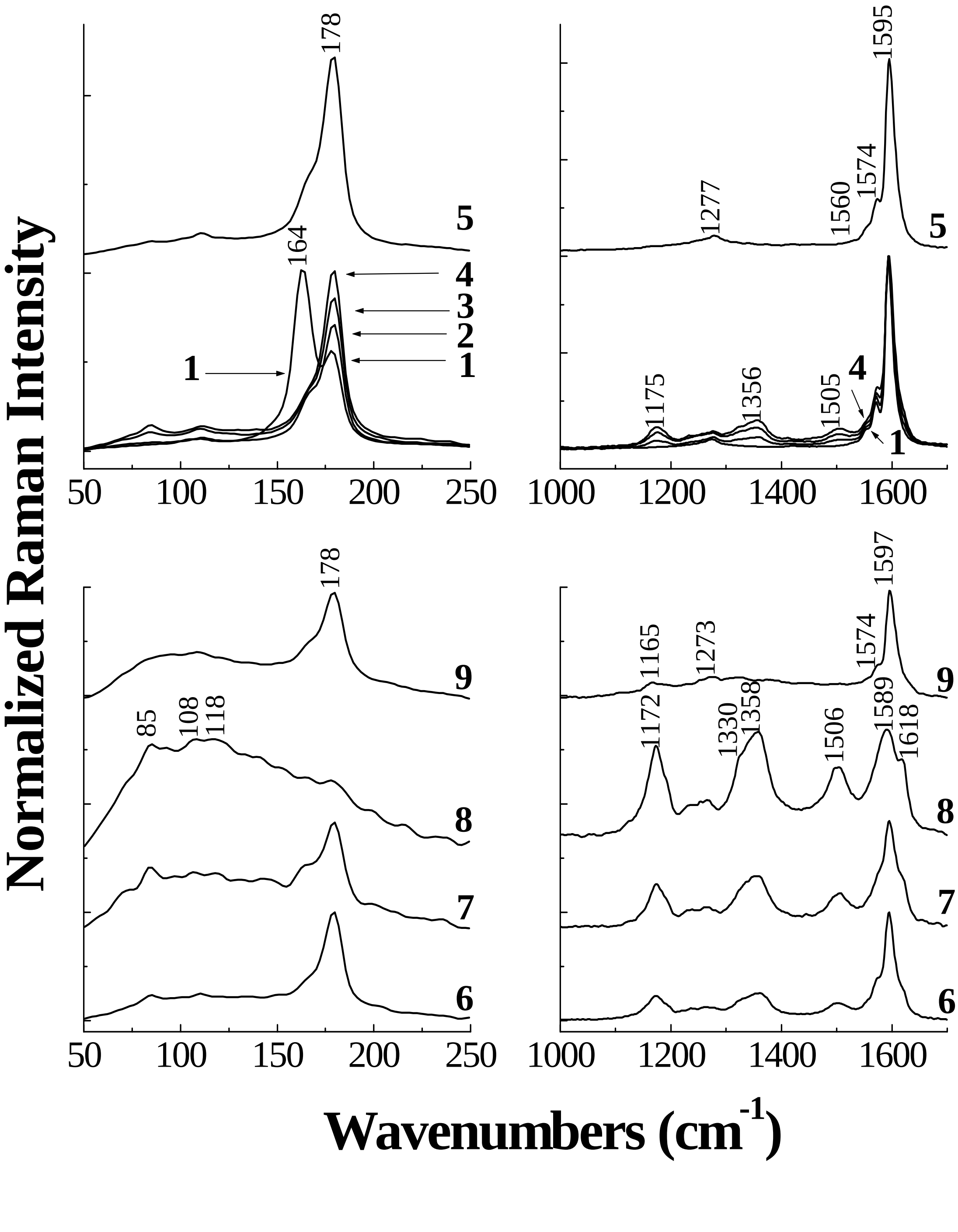

Fig. 2

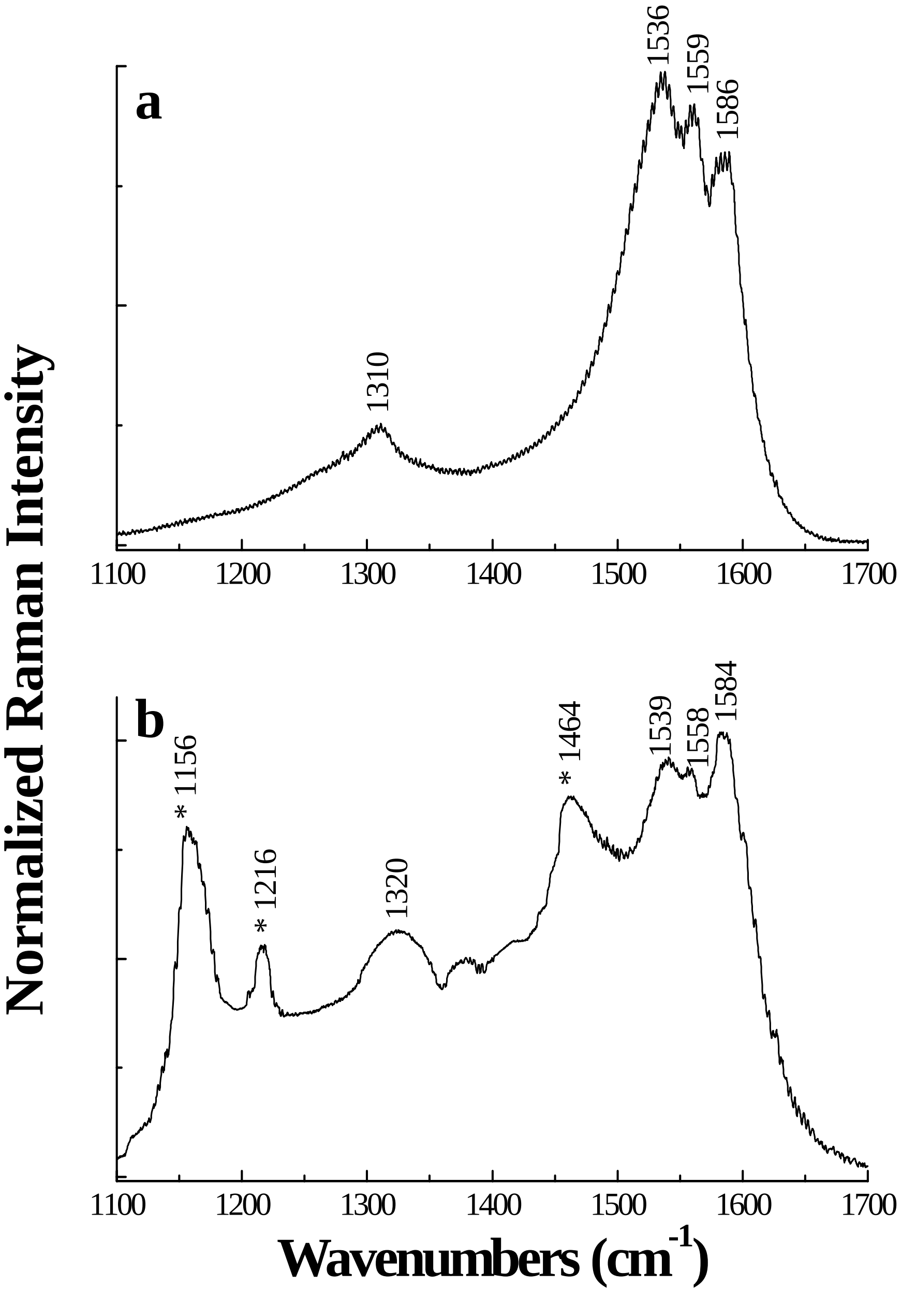

**Fig. 3**

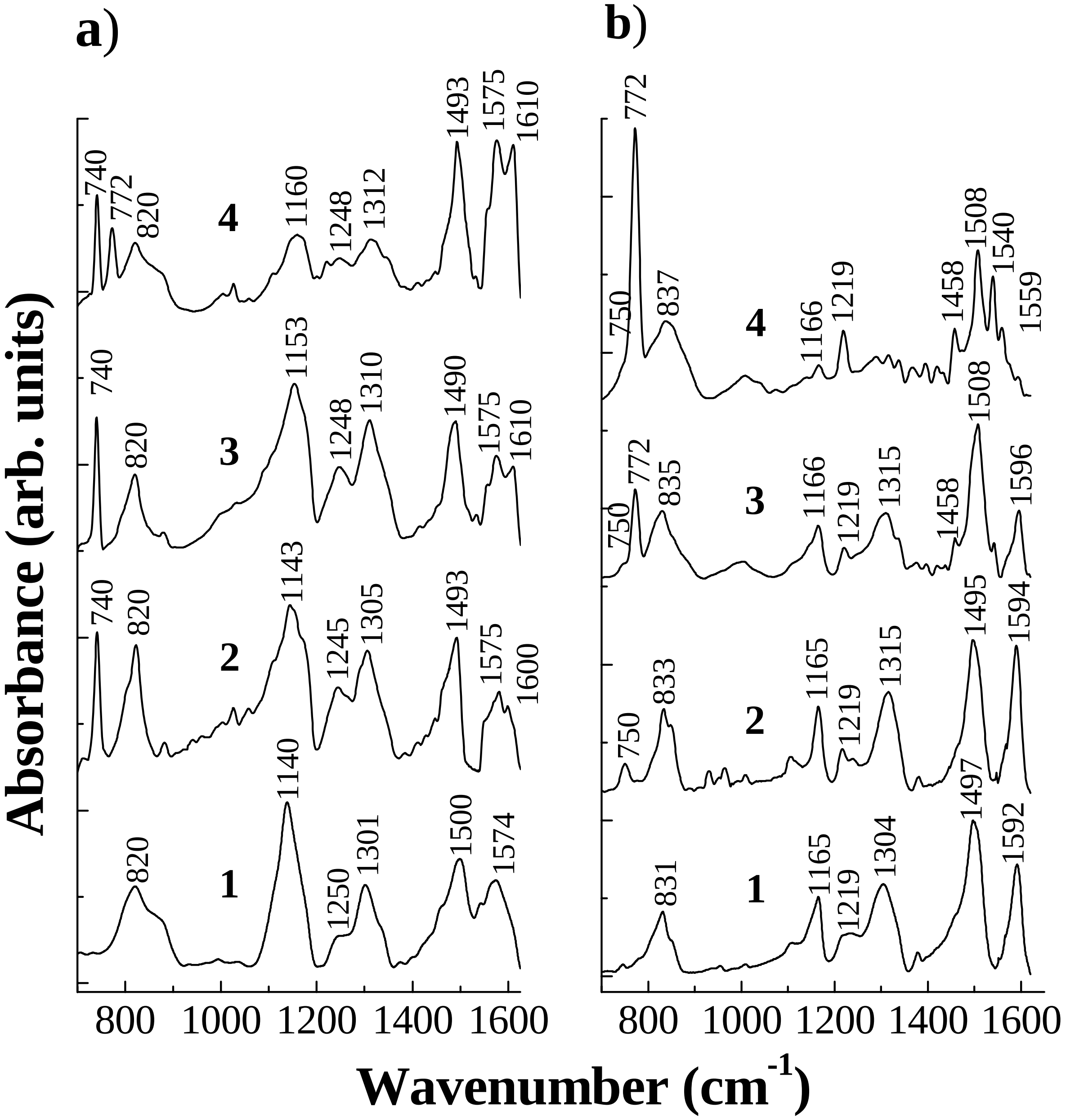

Fig. 4

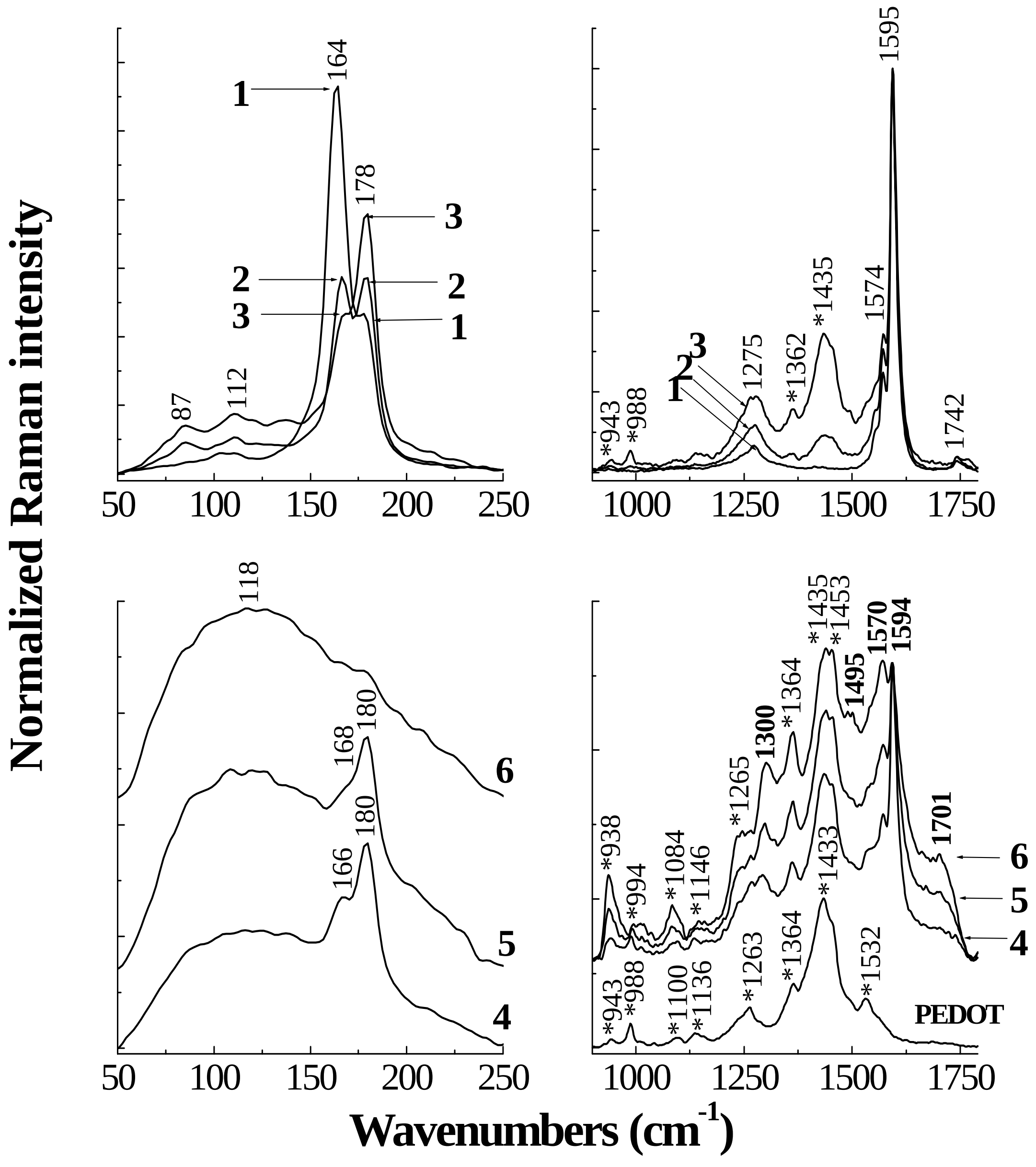

Fig. 5